\documentclass[%
superscriptaddress,
twocolumn,
nofootinbib,
 amsmath,amssymb,
 aps,
]{revtex4-2}

\usepackage{graphicx}
\usepackage{dcolumn}
\usepackage{bm}
\usepackage{color}

\definecolor{red}{rgb}{0.8, 0.0, 0.0}
\definecolor{blue}{rgb}{0.06, 0.2, 0.65}
\definecolor{green}{rgb}{0.0, 0.6, 0.0}
\usepackage{gensymb}
\usepackage{float}

\def\tpia#1{}

\begin{document}

\preprint{APS/123-QED}

\title{Nonuniform pressure helps structural superlubricity}

\author{Melisa M. Gianetti}
\affiliation{Institutt for maskinteknikk og produksjon, NTNU, Richard Birkelands vei 2B, 7034 Trondheim, Norway}
\author{Viet Hung Ho}
\affiliation{Institutt for maskinteknikk og produksjon, NTNU, Richard Birkelands vei 2B, 7034 Trondheim, Norway}
\author{Bj$\o$rn Haugen}
\affiliation{Institutt for maskinteknikk og produksjon, NTNU, Richard Birkelands vei 2B, 7034 Trondheim, Norway}
\author{Graham Cross}
\affiliation{School of Physics and CRANN, Trinity College Dublin, Dublin 2 D02 W085, Ireland}
\author{Astrid S. de Wijn}%
 \email{astrid.dewijn@ntnu.no}
\affiliation{%
Institutt for maskinteknikk og produksjon, NTNU, Richard Birkelands vei 2B, 7034 Trondheim, Norway}

\date{\today}

\begin{abstract}
Structural superlubricity, nearly vanishing friction between two structurally incommensurate crystalline surfaces, is a promising avenue for reducing friction in applications, but requires very specific and well-controlled conditions.
One of those conditions is perfectly uniform atomically flat surfaces.
Real-world surfaces are generally rough, leading to nonuniform pressure distributions.
We investigate the effects of nonuniform pressure distributions on structural superlubricity, using analytical calculations for rigid contacts as a basis, and molecular-dynamics simulations for a simple model to include the crucial effects of elasticity.
We show that a key ingredient is the vanishing pressure at the edge of the contact, and that this leads to improved scaling depinning and scaling behaviour, leading to lower friction.
We thus show that nonuniform pressure distributions actually help structural superlubricity, rather than hinder it.
\end{abstract}

\maketitle

\section{Introduction}

\tpia{This Paragraph Is About}

\tpia{Superlubricity is cool.}
\tpia{Superlubricity is potentially extremely useful.}
When two perfect crystalline surfaces with sufficiently different geometry are brought into contact, they can slide with vanishing friction.  This phenomenon is called structural superlubricity, and was first proposed theoretically by Shinjo and Hirano~\cite{Hirano_1990,Shinjo_1993}, and later demonstrated experimentally under various conditions~\cite{Martin_1993,Erdemir_2000,Dienwiebel_2004}.
While structural superlubricity has been suggested as a mechanism for low friction in a number of larger-scale systems, it has been shown definitively only in very controlled lab conditions, often in UHV, though sometimes in controlled nitrogen atmospheres or similar, and with very carefully engineered flat crystalline interfaces~\cite{Baykara_2018}.
In the experimental conditions where it has been demonstrated, the chemistry of the interfaces is crucial and the surfaces have to be atomically flat and sufficiently large.
\tpia{But it doesn't work in the real world yet, because ....}
Real surfaces in engineering conditions are not flat, but rough, leading to small contact asperities where the roughness peaks meet.
Consequently, we currently do not have any commercially usable engineering solutions that exploit this mechanism.

\tpia{Rough surfaces are curved, but is that the real problem?}
\tpia{What have other people done to look at this?}
It has been generally assumed that any roughness will destroy structural superlubricity.  Thus much experimental effort has gone towards creating larger and larger atomically flat surfaces (see for example~\cite{HanPRL_2026}).
One reason that flat surfaces are thought to be crucial, is that high local pressures can lead to damage to the surface, which destroys the crystalline order that is needed (see for example~\cite{Sun2024}).
However, not all surface curvature necessarily leads to such high pressures, and purely elastic deformation is also common in non-flat geometries with sufficiently gentle curvature.

\tpia{Curvature is interesting because there are reasons to think it will do something.}
Contacts between curved surfaces differ in a number of crucial qualitative ways from traditional flat superlubric contacts.
Firstly, the edges of contacts between curved surfaces are very different from the edges of flat contacts, and the edges of finite contacts play an important role in superlubricity (see for example~\cite{Sharp2016_PRB,deWijn2012,Hu2024,Wang2020}).
In addition, since the local stresses are no longer constant, there is a more complex interplay between the local stress and  elasticity of the materials.  Elastic deformation of the lattice can cause structural superlubricity to fail~\cite{Peyrard_1983}.

\tpia{Let's investigate!}
\tpia{What has been done in this work.}
In this work, we analyse the effect of a nonuniform load on structural superlubricity, and investigate the role of the edges and elasticity.
Specifically, we consider Hertz-like contacts between curved surfaces.
We perform analytical calculations for the scaling of friction with contact size in a rigid contact.
We show that the boundary conditions at the edge are crucial, and, rather than harming the structural superlubricity, actually improve it.
We then construct a simple model inspired by the Frenkel-Kontorova model~\cite{Frenkel_1938} with an elastic contact that allows us to numerically study the dependence of friction with the contact size, the total load and the elasticity through several orders of magnitude.
We show that the edges and elasticity together play a crucial role, and that elasticity also, up to a point, helps structural superlubricty.

\tpia{presentation of the first theoretical idea of the model}

\section{\label{sec:analytical_summation}Scaling in the rigid limit: analytical summation}

Structural superlubricity is characterised by a sublinear dependence of the friction on the contact area.
To investigate this in non-uniform contacts, we consider a rigid slider  with interactions constructed to represent a Hertz-like contact.
We investigate the scaling of the friction with contact area, which can be used to characterise the structural superlubricity~\cite{dietzelPRL2013,deWijn2012}.
To find the scaling exponents for the friction force with the contact area, we employ the analytical approach from Ref.~\cite{deWijn2012} and extend it to nonuniform contacts.

Atoms in the sliding body are subjected to a corrugated potential-energy landscape that is periodic with the lattice. To represent the nonuniform pressure distribution in a real non-flat contact, we modify this with a parabolic modulation. The potential energy of the sliding body $V (X)$ at the position $X$ on the surface can be written as a sum over the potentials of the $n$ atoms in the sliding body, indexed by $j$,
\begin{equation}
    V (X)  = \sum_{j} v ({X}_{j} , X) 
\end{equation}
Here $v (X_{j} , X )$ is the potential energy of an atom at position $X_{j}$ when the average position (center of mass) of the contacting atoms in the sliding object is $X = (1/n) \sum_{j} {{X}_{j}} $.

We split the potential energy into two factors, one periodic function due to the periodicity of the substrate, and one slowly-varying function $f$ representing the pressure profile. The function $f$ depends purely on the position inside the sliding object, so that we may write
\begin{equation}
    v ( {X}_{j}, X ) = \cos \left( \frac{2 \pi}{a} X_{j} \right) f  \left(X_{j}- X\right)~,
\end{equation}
where $a$ is the lattice parameter of the surface.
The function $f$ in principle depends on the elastic properties, shape of a contact asperity, adhesion between the surfaces, etc. Here, we consider a Hertz-like contact, which is a good approximation for a contact between two curved elastic surfaces.
While we do not know the exact relation between the local pressure and energy corrugation, we can assume that it is smooth, and thus $f$ has some kind of symmetric shape.  If the function is monotonic, the shape will be parabolic, but this is not required for our calculation.
We can then limit ourselves to well-behaved general functions of the form,
\begin{equation}
f  (x) =\begin{cases}
    V_0 \left[1-\sum_{l=1}^\infty C_l \left(\frac{x}{r}\right)^{2l} \right], & \text{if $|x| <r$}.\\
    0, & \text{otherwise}.
  \end{cases}
\end{equation}
Because the potential-energy landscape should be continuous in the edge points $|x| = r$, we have for the coefficients
\begin{equation}
\sum_{l=1}^\infty C_l = 1~.
\label{eq:sumC}
\end{equation}
This functional form will turn out to be tractable when summed over the atoms of the rigid crystal below.

The total potential energy of a one-dimensional crystalline contacts with Hertz-like corrugation can now be written as a sum over all $n$ atoms in the sliding body.
Let $b$ be the lattice spacing of the sliding body.
We can keep the expression as simple as possible by defining the sum such that we can write $X_j=X+bj$.
Thus,
\begin{widetext}
\begin{eqnarray}
V(X)
&=& \sum_{j=-(n-1)/2}^{(n-1)/2} \cos\left(\frac{2\pi}{a} (b j +X) \right) V_0 \left[1-\sum_{l=1}^\infty C_l \left(\frac{bj}{r}\right)^{2l}\right]~.
\label{eq:v1dbasic}
\end{eqnarray}
In order to evaluate this sum, we make use of the special sum $\sum_{k=0}^{n-1}\alpha^k=(1-\alpha^n)/(1-\alpha)$.  Using $\alpha=\exp(i\omega)$ this gives $\sum_{j=-(n-1)/2}^{(n-1)/2} \exp(i\omega j) = \sin[\omega n/2] /\sin(\omega/2)$.
Taking the $2l$-th order derivative with respect to $\omega$ on both sides gives us $\sum_{j=-(n-1)/2}^{(n-1)/2} (ij)^{2l} \exp(i\omega j) = (\partial^{2l}/\partial \omega^{2l} )[\sin[\omega n/2] /\sin(\omega/2)]$.

Splitting the $\cos$ in Eq.~(\ref{eq:v1dbasic}) into two complex exponentials, and using the above expressions to evaluate the sums, we obtain
\begin{eqnarray}
V(X) = \left. V_0 \cos\left( 2 \pi \frac{X}{a}\right) \left( 1 - \sum_{l=1}^\infty C_l \left(\frac{b}{r}\right)^{2l} (-1)^l \frac{\partial^{2l}}{\partial \omega^{2l}} \right) \frac{\sin(\omega n/2)}{\sin(\omega/2)}\right|_{\omega = 2 \pi b/a}~.
\label{eq:v1dintermediate}
\end{eqnarray}

To obtain the scaling of this expression with the number of particles $n$ in the sliding object, we expand the derivatives.
We also make use of the fact that $(n-1)/2< r < (n+1)/2$ to expand any prefactors containing $r$.
By evaluating the derivatives and keeping the leading order and next to leading order, and under the assumption that $b/a$ is not a half-integer,
we find  that
\begin{eqnarray}
V(X) &=& \left. V_0 \cos\left( 2 \pi \frac{X}{a}\right) \left\{ \frac{\sin(\omega n/2)}{\sin(\omega/2)} \right.\right.\nonumber\\ &&\left.\left.- \sum_{l=1}^\infty C_l \frac{1}{(n/2)^{2l} + O(n^{2l-1})} \left[\left(\left(\frac{n}{2}\right)^{2l} + O(n^{2l-1}) \right)   \frac{\sin(\omega n/2)}{\sin(\omega/2)} + O(n^{2l-1}) \cos(\omega n/2) \right]\right\} \right|_{\omega = 2 \pi b/a}
~.
\end{eqnarray}
\end{widetext}
From Eq.~(\ref{eq:sumC}), we see that the leading order cancels, and this gives
\begin{eqnarray}
V(X) = 
O(1/n) V_0 \cos\left( 2 \pi \frac{X}{a}\right)~.
\end{eqnarray}
The scaling exponent in this expression (-1) is smaller than the exponent of 0 found for a 1-d incommensurate interface with uniform corrugation~\cite{deWijn2012}.
This sublinear scaling is indicative of structural superlubricty, but also demonstrates that nonuniform contacts without any in-plane elasticity could have even lower friction than uniform ones.
It is worth noting that the fact that the leading order cancels is a direct consequence of the corrugation amplitude vanishing at the edge in a continuous way.  This is qualitatively different from a flat contact with a uniform pressure distribution, where there is a jump in the corrugation at the edge and the leading order term does not vanish.
This is consistent with the picture that in incommensurate interfaces it is the edge that dominates the total friction.

The analytical calculations thus show that for this general class of nonuniform corrugation, the scaling exponent of the total corrugation with the contact area is at most $-1$.
This is actually a lower exponent than in the case of uniform corrugation~\cite{deWijn2012}, which has a $0$ exponent.
We conclude from this that a relatively smooth but nonuniform pressure distribution is not a fundamental barrier to structural superlubricity.
Nevertheless, in order to keep this calculation tractible, it has been extremely idealized.  Elasticity and dynamical effects could interact with the nonuniform pressure distribution in nontrivial ways and affect the superlubricity.

\section{\label{sec:methods}Elastic model and simulation setup}

Even in uniform contacts, the robustness of low-friction states in incommensurate contacts is fundamentally limited by the interplay between energy corrugation and bulk elasticity~\cite{Peyrard_1983}, characterised by the Aubry transition.  If the material is too soft, or the corrugation (or load) too high in comparison, the positions of the atoms in the contact are controlled by the surface energy, rather than the lattice of the counter surface, leading to high friction.
In a contact with a nonuniform pressure distribution, there are nonuniform stresses in the slider, and consequently nonuniform elastic deformation, which in turn may affect this transition.

To investigate the effect of elasticity, we perform molecular-dynamics simulations of a more realistic model system, which includes an elastic slider inspired by the Frenkel-Kontorova model~\cite{Frenkel_1938}, combined with surface-chain interactions set up in a way that allows for the application of a nonuniform pressure in the contact.
In order to capture the scaling of friction with size, we need a size range over several orders of magnitude.
For reasons of computational efficiency, we thus restrict ourselves to a two-dimensional system with a chain on a one-dimensional surface.

\subsection{\label{sec:model}Model details}

\begin{figure}
    \centering
    \includegraphics[width=1\columnwidth]{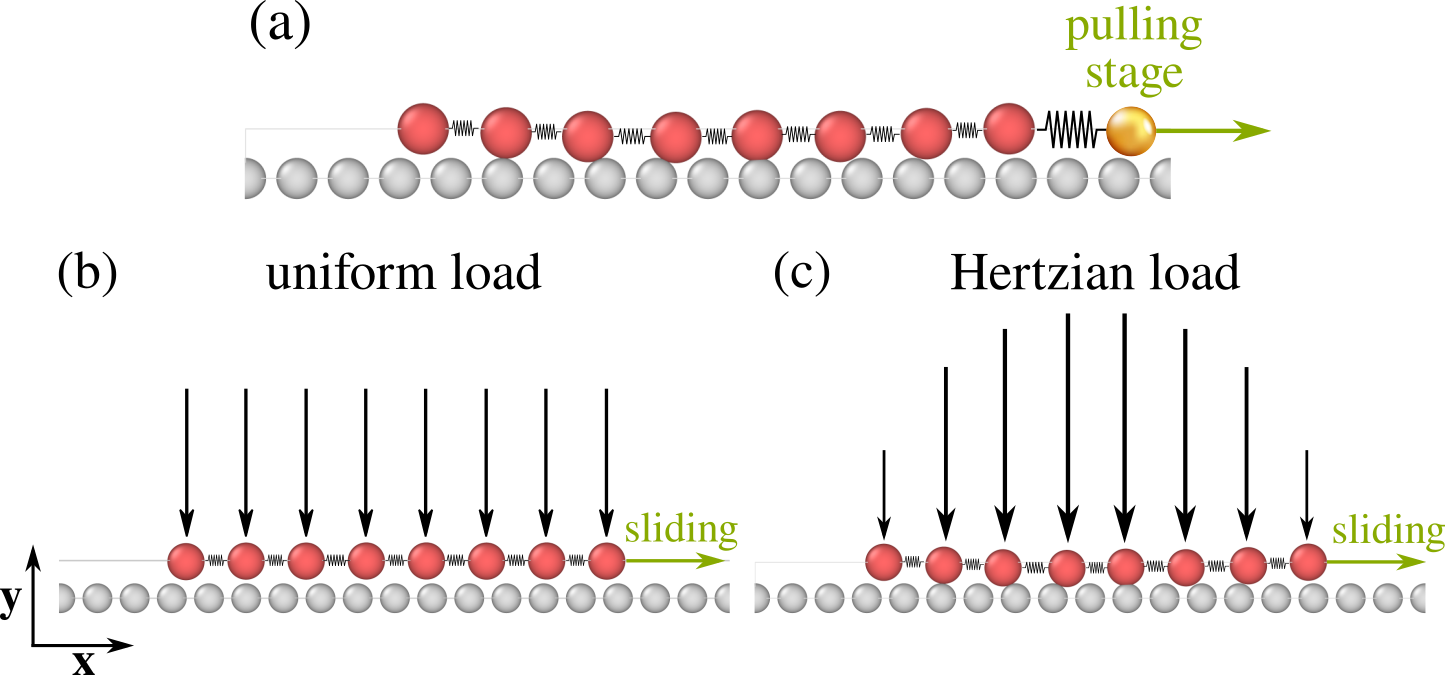}
    \caption{(a) Sketch of the model for a slider of $n=8$ particles. (b) Uniform and (c) Hertzian loads applied to the system.}
    \label{fig:system}
\end{figure}

Like in the Frenkel-Kontorova model, our system consists of a harmonic elastic chain of particles.
However, rather than keeping the chain to a simple period potential, we place it on a rigid crystalline substrate and subject it to a pressure distribution.
We construct our model system in 2D, with a slider and a surface which is periodic in the $x$ direction in order to mimic an infinite substrate.
All quantities are expressed in reduced units.
Figure~\ref{fig:system}a shows a sketch of the system for an 8-particle slider. 

The surface particles are arranged in a regular lattice with lattice parameter $a$.  All other lengths will be expressed in terms of $a$.
The slider consists of a finite chain of $n$ particles interacting through harmonic bond and angle potentials.
In order to obtain a convenient range of slider sizes, we have chosen $n$ following the Fibonacci series starting from $n = 21$ to 10946.
For a distance $r$ between two nearest neighbors in the chain, the potential energy is given by
\begin{equation}
	V_{\mathrm{bond}}(r)=k_{\mathrm b} \left (r - b \right)^{2}~,
    \label{eq:bonds}
\end{equation}
with stiffness $k_{\mathrm b} = 500 \epsilon/a^2$ (with $\epsilon$ the characteristic energy of the surface-chain particle interaction) unless otherwise specified, and an equilibrium bond length $b/a=1$ for commensurate configurations and $b/a = \frac{1+\sqrt{5}}{2}$ for incommensurate configurations. 
The bending energy of the bonds in the chain is a function of the angle $\theta$ between two subsequent bonds in the chain, which gives an energy
\begin{equation}
	V_{\mathrm{angle}}(\theta)=k_{\mathrm \theta} \left (\theta - \theta_\mathrm0 \right)^{2}~,
    \label{eq:angles}
\end{equation}
with a stiffness $k_{\mathrm \theta} = 500 \epsilon/\mathrm{rad}^2$ and an equilibrium angle $\theta_\mathrm 0 = \pi$.
The first and last particle in the chain have no neighbor on one side, as they are at the edge of the finite-size Hertz contact.

The slider and surface interact through the standard (12,6) Lennard-Jones (LJ) potential, defined for the interaction between atoms $i$ and $j$ as:
\begin{equation}
	V_\mathrm{LJ} (r)=4\epsilon \left[\left(\frac{\sigma}{r}\right)^{12}-\left(\frac{\sigma}{r}\right)^{6}\right], r < r_\mathrm{c}
    \label{eq:LJpotetnial}
\end{equation}
where $r$ is the distance between the interacting particles, $\epsilon$ is the potential well-depth and $\sigma=a$ is the distance at which the interparticle potential is zero.
In our simulations, the global cutoff $r_\mathrm{c} =2.5\sigma$ was applied to all interactions.
All quantities in the rest of the manuscript will be expressed in terms of the LJ interaction parameters $\epsilon$ and $\sigma=a$.  To complete the unit system, the mass of the slider particles is also set to unity, so that the time unit $\tau=1\sqrt{\epsilon/(ma^2)}$.

The slider is attached by its center of mass to a pulling stage using a harmonic spring that acts only on $x$.
The pulling stage moves at constant velocity parallel to the surface.
The spring constant scales linearly with the number of particles in the slider, $k = (n/10) \epsilon/a$.

We apply two different pressure profiles, as sketched in Figures~\ref{fig:system}b and~\ref{fig:system}c: a uniform load where all the slider particles are subjected to the same amount of load, and a Hertzian pressure profile where the load depends on the location of the particle in the slider.
When comparing the cases of uniform and Hertzian loads, we compare two cases with the same total pressure, i.e.\ the same average load per particle.
The Hertzian pressure profile is parabolic with a maximum in the center of the slider and vanishing pressure at the edges.
It is given by
\begin{equation}
l_{j}\propto\sqrt{1-\left\{  4 \cdot \left[ \frac{X_{j}}{\left(n-1  \right)\cdot b}   \right]^{2} \right\}}~,
\label{eq:hertz_load}
\end{equation}
where $n$ is the number of particles in the slider, $b$ is the equilibrium distance between the particles in the slider, $X_{j}$ is the equilibrium position of the $j$th particle in the slider in the reference frame of the center of mass of the slider.
This distribution gives a vanishing pressure for the edge particles, and maximum pressure in the center.
We refer to this nonuniform pressure as the Hertzian pressure profile in the rest of this manuscript.

\subsection{\label{sec:simulations}Simulation details}
The simulations are carried out with LAMMPS~\footnote{www.lammps.org}~\cite{lammps2}. 
The slider was constructed using Moltemplate~\cite{moltemplate}.

The slider is first equilibrated without load close to the surface using a three-stage minimization to ensure a stable, low-energy contact state before applying the vertical load to each particle. The choice of this protocol aims to equilibrate both small and large sliders. In the first stage, a steepest descent (SD) algorithm was employed with a limited displacement per step of $d_{\mathrm{max}}=0.01\sigma$ to eliminate high-energy steric overlaps and relax initial bond and angle stresses without numerical instability.
Subsequently, a refining stage using the semi-dynamical fire inertial relaxation engine (FIRE) was used to accelerate the convergence towards the local potential energy minimum with a force tolerance of $10^{-8}\epsilon/a$ . 
Finally, we perform a dynamical quench with viscous damping time $0.5\tau$ for a time of
$10\tau$. This final stage dissipates residual kinetic energy and allows the long-wavelength vibrational modes in the flexible slider to settle, ensuring that it has reached its equilibrium separation distance from the rigid surface.  

Once we have a minimized configuration for the slider sitting on the surface, the vertical load is imposed and the slider is coupled to the pulling stage located directly in front of the first atom of the slider, what we call the ``head atom''.  The puller starts moving immediately with a speed of 0.1$a/\tau$, and in most cases the slider also starts moving immediately as a result.  Since we are interested in the steady-state sliding, and not static friction, this starting procedure is designed to produce a steady state quickly for all contact sizes.  It does not affect the final steady state, as long as the slider is sufficiently close to the surface at the beginning.
The instantaneous friction force is provided by the force in the puller spring, which is proportional to its elongation. The friction is then calculated from the average once the system has reached a steady state. 

Energy is removed from the system through a viscous damping with damping parameter $\gamma_{\mathrm{iy}}=\gamma_{\mathrm{iz}}=\gamma=0.05$\,$\tau$, ensuring that the system is strongly damped and operates in the quasi-static regime. The time step is $0.005\tau$.
We run simulations with values for the load per particle from 1 to 1000 $\epsilon/a$.
As we are interested in the static behaviour, the average kinetic force due to the damping ($nm v/\gamma=2 nma/\tau$) is subtracted.

\section{\label{sec:results}Results and discussion}

\begin{figure*}
    \centering
    \includegraphics[width=0.8\textwidth]{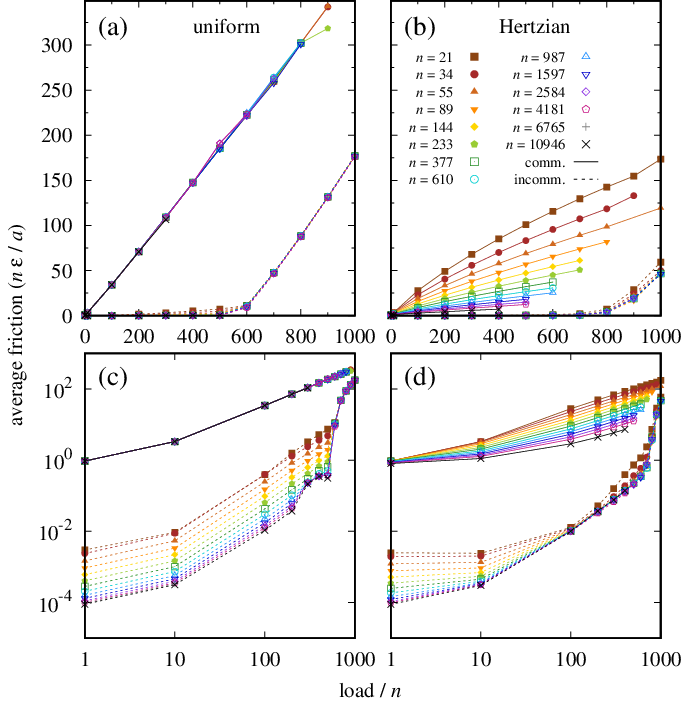}
    \caption{Average friction per particle as a function of the load per particle for commensurate and incommensurate systems with (a) and (c) uniform and (b) and (d) Hertzian pressure distributions. (a) and (b) linear scale, (c) and (d) log-log scale. 
    Solid lines represent the commensurate contacts, and dashed lines represent incommensurate contacts.  Different colors represent different slider sizes ($n$). The incommensurate system has lower friction than the commensurate for both uniform and nonuniform pressure distributions.
    }
    \label{fig:UvsH}
\end{figure*}

\tpia{We start by analyzing the uniform pressure distribution.}
As a baseline, we first focus on the case of uniform pressure.  This has been well-studied in both commensurate and incommensurate contacts.
\tpia{Commensurate contact with uniform load: expected behavior.}
For a commensurate contact under uniform pressure the average friction per slider particle is generally expected to be linear in the load per particle and independent of the size of the slider. Figures~\ref{fig:UvsH}a and~c show this behavior. 
\tpia{incommensurate contact with uniform load: also expected behavior and Aubry transition}
We also see in the figure that in the incommensurate case the friction is lower than in the commensurate case for all sizes and loads per particle, as also expected.
In addition, at low loads, the friction barely increases, until a transition occurs at around $600\epsilon/a$. 
This is the Aubry transition~\cite{Peyrard_1983}, which occurs when the forces due to the interaction with the surface start to dominate over the springs between the particles in the chain. The position of the particles is then controlled by the surface, no longer by the springs in the chain or their incommensurate equilibrium spacing, destroying the incommensurability and causing particles to become pinned.  At the Aubry transition, the friction thus qualitatively changes to be linear with the load, matching the commensurate slope, only with an offset.
Below the Aubry transition, there is already a signature of the approaching transition in a powerlaw behaviour that is related to long-range relaxation.

\tpia{Nonuniform load: key difference with uniform and what does this suggest? That SL is maintained for higher loads.}
When a nonuniform pressure distribution is applied (Figures~\ref{fig:UvsH}b and~d), the picture changes.
The friction is still smaller in the incommensurate case than in the commensurate case.
However, for both commensurate and incommensurate systems, the nonuniform distribution results in lower overall friction compared to the uniform case.
Commensurate contacts under Herzian pressure distribution display lower friction with sublinear dependence, due to dislocations moving under stress gradients~\cite{Sharp_2017}, as we will show in more detail below.
The nonuniform pressure also smears out the Aubry transition observed for incommensurate sliders with uniform pressure; rather than an abrupt jump, we observe a more gradual onset of pinning.
\tpia{Commensurate nonuniform load brief explanation}

\tpia{closer look to the data of Fig \ref{fig:UvsH} a and b: dependence of the friction with the contact size at low loads}
Another signature of the Aubry transition can be seen in Figs.~\ref{fig:UvsH}c and~\ref{fig:UvsH}d in the change in the dependence on contact size.
For the commensurate contact under uniform pressure, there is no dependence of the friction coefficient with the contact size.
For the incommensurate case, the contact-size dependence is strong at lower loads.
At higher loads, after the Aubry transition has set in, it behaves more similarly to the commensurate contacts.

\begin{figure}
    \centering
    \includegraphics[width=1\columnwidth]{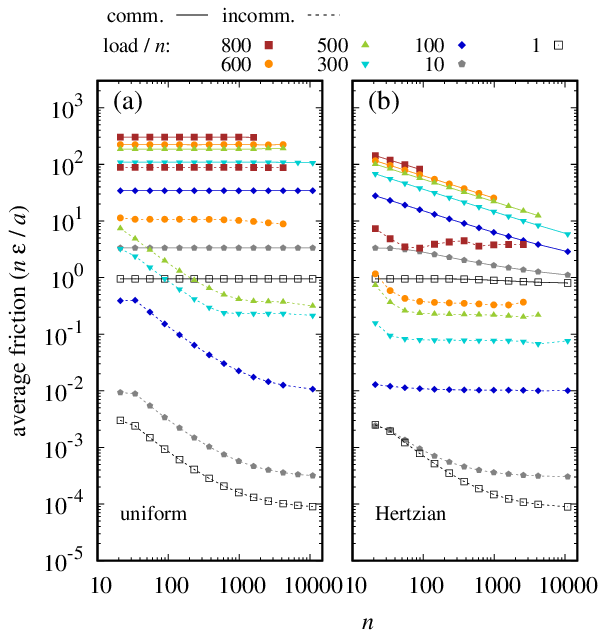}
    \caption{Average friction per particle as a function of the number of particles, log-log plot for both commensurate and incommensurate systems with (a) uniform and (b) Hertzian pressure distribution. Solid and dashed lines correspond to commensurate and incommensurate contacts respectively.  The observation of figure~\ref{fig:UvsH} is confirmed. Both systems show lower friction when the distribution of the pressure is non uniform.}
    \label{fig:UvsH_loglog}
\end{figure}

\tpia{analysis of the scaling of friction on Fig \ref{fig:UvsH_loglog}: uniform load}
To elucidate the dependence on the contact size further, and because size scaling is an indicator of structural superlubricity, we show the friction as a function of contact size in Figure~\ref{fig:UvsH_loglog}. 
As before, both systems show lower friction when the distribution of the pressure is nonuniform. For the commensurate system with uniform pressure (Figure~\ref{fig:UvsH_loglog}a, solid lines), the friction per particle does not depend on $n$, as expected, indicating that the friction scales linearly with the contact size. For the incommensurate system with uniform  pressure (Figure~\ref{fig:UvsH_loglog}a, dashed lines), the friction per particle decreases with the contact size, and the friction thus scales sublinearly with the contact size, as expected~\cite{dietzelPRL2013,deWijn2012,Sharp2016_PRB}, with the expected scaling exponent of -1 for the friction per particle (0 for the friction itself), up to large contact sizes. For large contacts, the long-range elasticity takes over~\cite{Mueser_2004,Sharp2016_PRB}, and long-range relaxation limits the incommensurability.

\tpia{analysis of the scaling of friction on Fig \ref{fig:UvsH_loglog}: nonuniform load}
Under nonuniform pressure (Figure~\ref{fig:UvsH_loglog}b), the commensurate system (solid lines) already displays a sublinear size dependence due to the dislocation mechanism found by Sharp et al~\cite{Sharp_2017}, as mentioned above, with a scaling exponent of approximately $-0.37$.
However, the exponent cannot be compared to Sharp et al.'s result directly, due to differences in dimensionality and setup.

For incommensurate contacts under nonuniform pressure (Figure~\ref{fig:UvsH_loglog}b, dashed lines), we have a combination of the effects of incommensurability and nonuniform pressure.
The friction is lower than both the nonuniform commensurate and uniform incommensurate cases.
The long-range elastic relaxation sets in more quickly at high loads than it does for the uniform pressure distribution, so that we do not have enough orders of magnitude to extract a scaling exponent.  Nevertheless, it seems unlikely that it would be the $-2$ predicted for rigid crystals in Sec.~\ref{sec:analytical_summation}
Nevertheless, for all cases the friction for the nonuniform pressure is lower than, or at the worst the same as for, the uniform case.

\begin{figure*}
    \centering
    \includegraphics[width=0.9\textwidth]{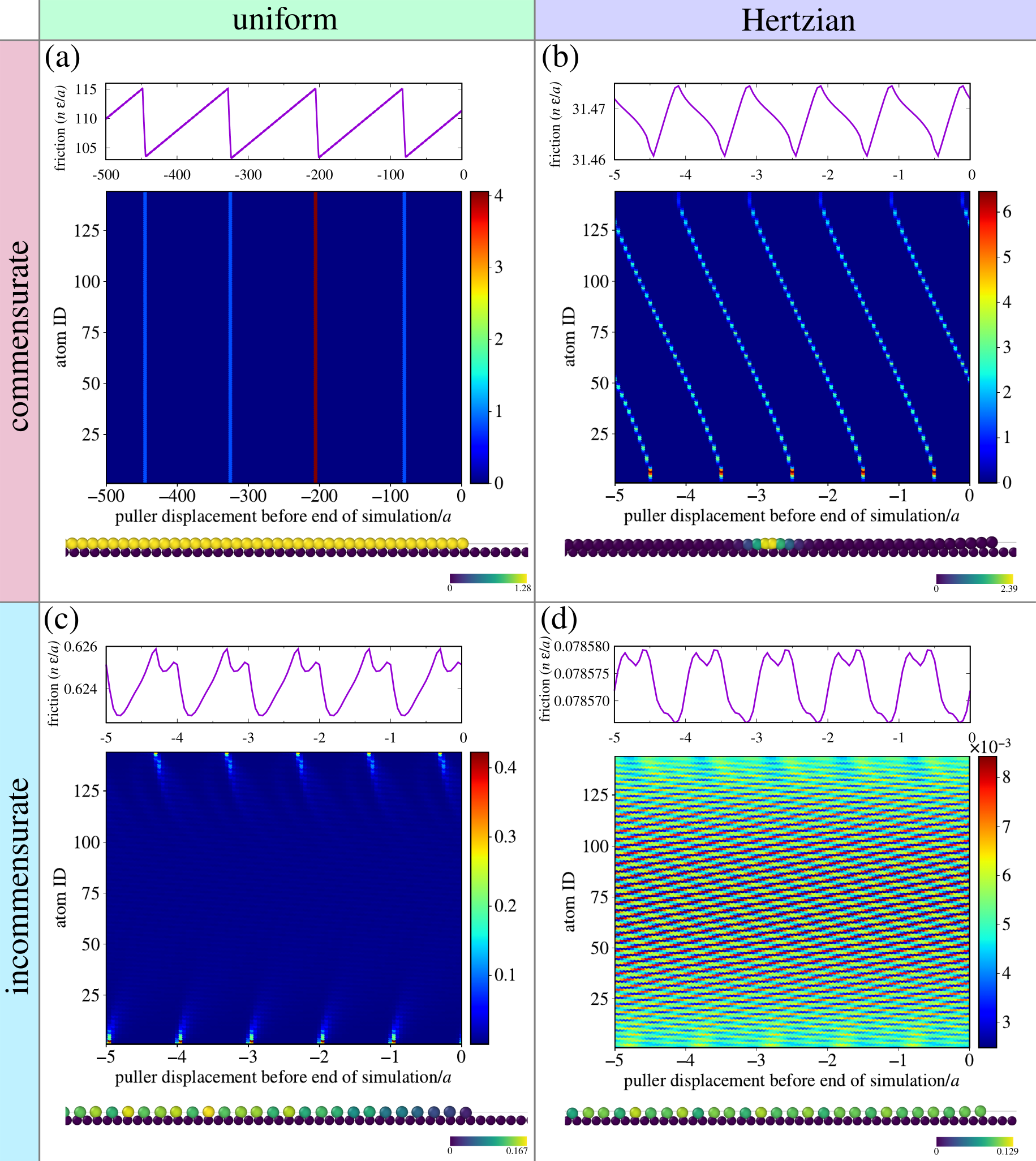}
    \caption{Kinetic energy maps at load$/n = 300 \epsilon/a$ (below the Aubry transition), $n = 144$ for a small displacement during the steady state, (a) commensurate configuration - uniform pressure, 
    (b) commensurate configuration - Hertzian pressure, 
    (c) incommensurate configuration - uniform pressure, 
    (d) incommensurate configuration - Hertzian pressure.
    On each panel on top, the friction per particle as a function of the displacement reported on the kinetic energy map, on the center the kinetic energy map, on the bottom a snapshot of the head of the slider with each particle colored by its velocity magnitude. Atom ID = 1 is the first atom of the slider, what we call the ``head atom''.
    }
    \label{fig:kinetic_energy_maps}
\end{figure*}

\subsection{Mechanics of sliding}

\tpia{we analyze the dynamics to explain the lower friction in nonuniform loads}
To unravel why the friction is lower for a nonuniform distribution compared to a uniform distribution with the same average load per particle, we investigate the mechanics of sliding in more detail.  Our system is strongly damped and in the quasistatic regime.
However, since the velocity of the slider is nonzero, we can still use the velocity of the particles as a tool to investigate the quasi-static mechanisms by which sliding occurs.

\tpia{typical examples below the Aubry transition shown in Fig \ref{fig:kinetic_energy_maps}}
In  Figure~\ref{fig:kinetic_energy_maps} we show some typical examples to illustrate the different behaviors.
It displays the kinetic energy of each particle in a slider of $n = 144$ as a function of the puller displacement for different pressure distributions for both commensurate and incommensurate configurations at $\mathrm{load}/n = 300$, which is well below the Aubry transition.
Each panel also contains a snapshot of the head of the slider at the bottom with each particle colored by the magnitude of its velocity.
The corresponding friction trace is shown at the top.
Atom ID = 1 is the first atom of the slider, what we call the ``head atom''.

\tpia{This is the order we do things in and why we do it like that.}

\subsubsection{Reference cases: commensurate}
\tpia{first case: uniform - commensurate}
We once again start from the simplest and most well-understood case as a reference, the commensurate contact with uniform pressure. 
The commensurate system with uniform pressure distribution (Figure~\ref{fig:kinetic_energy_maps}a) shows a clear stick-slip behavior as seen on the friction trace, when the slider is either stuck entirely, or moves at once when the force in the pulling spring overcomes the threshold of the corrugation (see Movie 1 of the Supplemental material~\cite{Supplemental}). The energy map shows vertical lines of movement in the entire slider at every slip, and the snapshot on the bottom of the panel, taken during a slip event, shows that all the slider particles have the same velocity and move together.

\tpia{second case: nonuniform - commensurate}
For the nonuniform load (Figure~\ref{fig:kinetic_energy_maps}b), instead waves of high kinetic energy move through the slider as it advances, starting from the tail end: the particles within the slider move in turns and not all at once (see Movie 2 of the Supplemental material~\cite{Supplemental}).
This is due to dislocations that nucleate at the edge and then move through the interface under the local stress gradient, similar to what is described in Ref.~\cite{Sharp_2017}.
This nucleation is possible due to the tapering of the load at the edges, so that the edge atoms can move much more easily than the bulk atoms.
When the dislocations move through the slider, energy is dissipated.
This is confirmed on the snapshot on the bottom of panel b of Figure~\ref{fig:kinetic_energy_maps}, where one can see the dislocation with particles at slightly higher positions and with non-zero velocity, while all the rest of the slider particles are pinned and closer to the surface (i.e. they have zero velocity like the surface).
On the friction trace we see a distorted stick-slip.  The peaks correspond to the creation of a dislocation, while the troughs correspond to the destruction of one.
\tpia{partial conclusion for the friction in the commensurate contact (the friction is HIGH!)}

The commensurate contact geometry allows the slider to be partially (Figure~\ref{fig:kinetic_energy_maps}b) or totally (Figure~\ref{fig:kinetic_energy_maps}a) pinned for long periods during the sliding, leading to high friction.
However, the nonuniform pressure forces different particles in the chain to slide at different times. When one particle moves, its neighbors, which would not have moved collectively yet, can move more easily. This triggers the dislocations traveling through the chain, rather than forcing a large-scale collective behaviour and leads to reduced friction.

\subsubsection{Reference case: incommensurate, uniform pressure}
\tpia{third case: uniform - incommensurate}
We now turn to incommensurate contacts, where the mechanisms are less directly obvious.
For uniform pressure distribution (Figure~\ref{fig:kinetic_energy_maps}c) the friction trace displays signatures of stick-slip, but is clearly distorted. The peaks in the forces are directly followed by steep drops accompanied by high kinetic energy near the edges.
At each slip, a wave of displacement thus moves relatively rapidly from the edges into the contact.  We also see that these slips do not occur for both ends at the same time (see Movie 3 of the Supplemental material~\cite{Supplemental}).
Away from the edges, the contact is not pinned, and particles move more steadily at constant, but lower, speeds.
The contact moves by sticking and slipping of the extremes (atomID 1 and 144 in this example). 
In general, the edges of finite-size incommensurate contacts are known to play a key role in their friction, for example for the scaling with the contact size~\cite{deWijn2012}, and this behavior is consistent with that picture.

\subsubsection{Incommensurate contact under nonuniform pressure}
\tpia{fourth case: nonuniform - incommensurate}
Now that we have established the phenomenology in the reference cases, we can use this to better understand the behavior of the incommensurate contact under nonuniform pressure
(Figure~\ref{fig:kinetic_energy_maps}d). The friction trace is qualitatively very similar to the one with the uniform pressure distribution but the variation of the friction force is much smaller. This is because the entire contact is constantly moving, as can be seen from the heat map of the kinetic energy.

This behavior is the result of a combination of two things: depinning of the edges due to the pressure distribution, and easy sliding in the incommensurate bulk.
Similar to the commensurate contact under nonuniform pressure, at the edges waves of displacement are triggered, and appear more readily than in the case of uniform pressure.
It is again the tapering of the load at the edges in the nonuniform incommensurate contact that prevents the pinning of the edges that we do observe for the uniform pressure distribution.
Combined with the depinning of the incommensurate bulk, this leads to a contact without any pinning at all.
All particles move with similar velocity (see the snapshot on the bottom of Figure~\ref{fig:kinetic_energy_maps}d),
with an overall smooth sliding motion (see Movie 4 of the Supplemental material~\cite{Supplemental}).
The results is an even lower friction than for the incommensurate contact with uniform pressure. 
\tpia{partial conclusion for the friction in the incommensurate contact (the friction is LOW!)}

\subsection{Elasticity}

The elasticity of the slider can play a nontrivial role for a number of reasons.
The stresses induced by the pressure gradient (i.e.\ different forces on neighboring particles in the slider) may compete with the stiffness of the springs in the slider.
The internal stiffness of the slider also serves as the primary resistance to substrate-induced pinning below the Aubry transition, which in turn is related also to the load.
It is therefore worth investigating the effect of the spring stiffness in the non-uniform incommensurate contacts, and how it interacts with the load.
We performed additional, analogous, simulations for higher and lower values than our default of $k_\mathrm{b}=500\epsilon/a$ to analyze this effect, $k_{\mathrm b} = 250, 1000, 2000\epsilon/a$.

\begin{figure*}
    \centering
    \includegraphics[width=\textwidth]{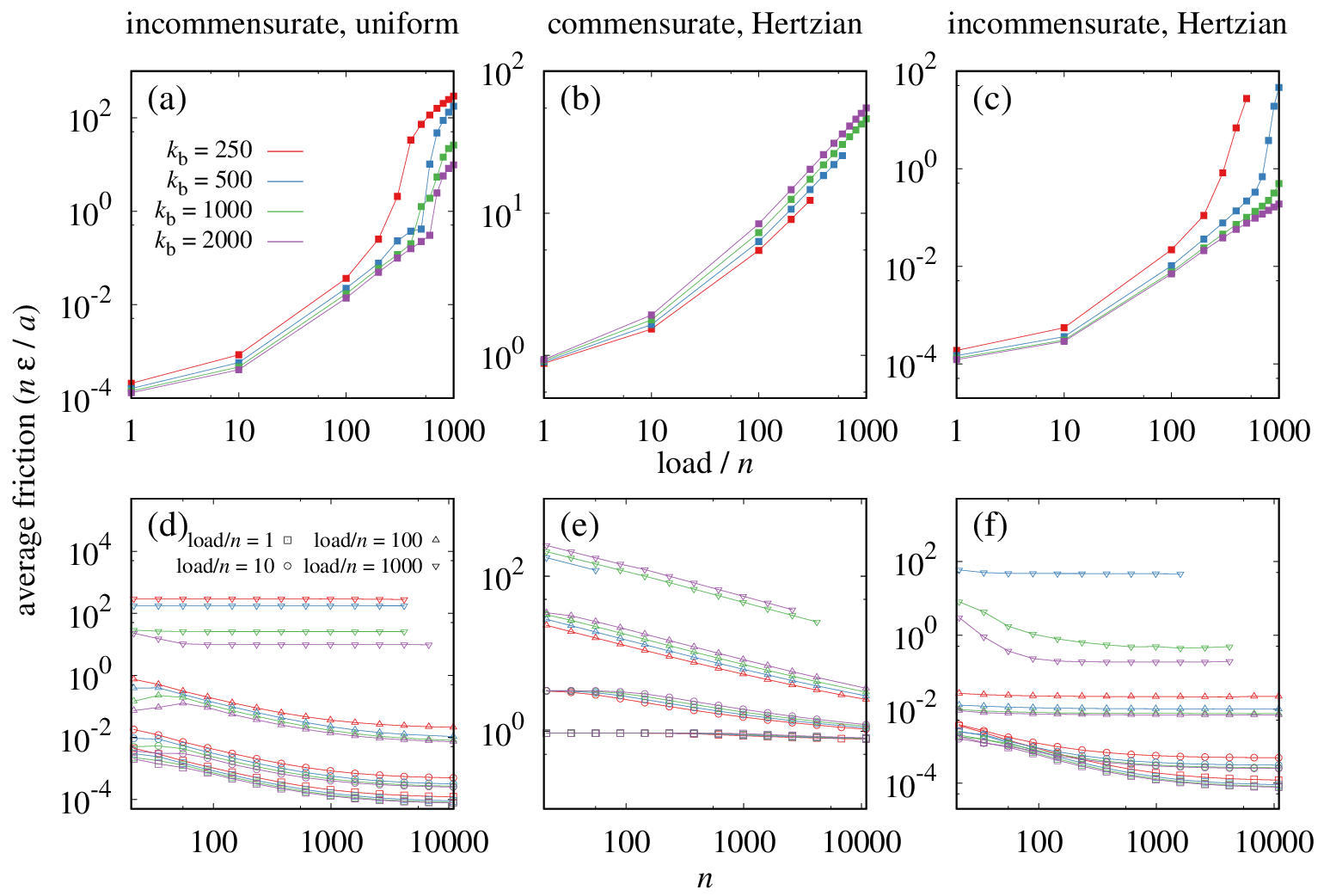}
    \caption{Elasticity effect on commensurate and incommensurate contacts under uniform and nonuniform load. Average friction per particle as a function of load per particle for (a) incommensurate, uniform load, (b) commensurate, nonuniform load and (c) incommensurate, nonuniform load for different bond elasticities. Average friction per particle as a function of the number of particles, log-log plot for (d) incommensurate, uniform load, (e) commensurate, nonuniform load and (f) incommensurate, nonuniform load for different bond elasticities.
    \tpia{\bf Stiffer sliders give lower friction in the incommensurate system and higher friction in the commensurate.}
    }
    \label{fig:elasticity}
\end{figure*}

Once again, for reference, we first consider the case of uniform load. 
In the commensurate case, there is no effect of the elastic constants, as all particles in the slider move together.
This is different for the incommensurate contact.
Figure~\ref{fig:elasticity}a shows the average friction per particle as a function of the load per particle and contact size for all four cases for the incommensurate contacts. 
We see that, as expected, the Aubry transition shifts to higher loads when the slider becomes stiffer (Fig.~\ref{fig:elasticity}a).
There is one load where the frictions for $k_\mathrm{b}=1000$ and $k_\mathrm{b}=500$ are reversed.  A detailed investigation of this beyond the scope of this work, but we have observed that there is a size dependence and we suspect it is related to the size of Moir\'e superlattices compared to the size of the slider, which can lead to nontrivial behavior~\cite{Huang2022}.

In Fig.~\ref{fig:elasticity}d, we also see the expected effect of elasticity and contact size. The superlubric sublinear scaling of the friction with contact size levels off earlier and to a higher value for the softer sliders. This is due to the aforementioned long-range elastic relaxation, the length scale of which becomes shorter as the springs become softer.
At the largest load of $1000\epsilon/a$, which is above the Aubry transition for all of these spring stiffnesses, the sublinear scaling disappears.

We now turn to the nonuniform pressure distribution. Figures~\ref{fig:elasticity}b and~e show the average friction per particle as a function of the load per particle and contact size for the same four cases, for the commensurate interface.
The stiffness of the slider affects the dislocations.
We see that the friction is lower for sliders with lower elastic constants.  This is because, in the commensurate regime, the stiffer the slider, the higher the Peierls-Nabarro barrier that must be overcome to move the dislocation, and thus the higher the friction.

Finally, the case of the nonuniform pressure and incommensurate contact is shown in Figures~\ref{fig:elasticity}c and~f.
As in the uniform pressure case, the Aubry transition is also shifted to higher loads for stiffer sliders.
The size scaling shows a similar trend as in the uniform pressure case: the softer sliders have higher friction that levels off earlier, as expected from the long-range relaxation mechanism with softer springs.
However, this effect is more pronounced, due to the extra reduction in friction from the depinning at the edges of the contact.
In the incommensurate regime, below the Aubry transition, the Peierls-Nabarro barrier decreases with increasing stiffness, making depinning easier.
Thus, the increased stiffness leads to lower friction here through both the edge- and Aubry-related mechanisms.

\section{Conclusions}
We have investigated superlubricity in contacts with nonuniform pressure distribution, which is more common in realistic contacts than uniform pressure.  We focus on contacts with a Hertz-like pressure profile.
We have performed analytical calculations for rigid contacts, and numerical simulations to investigate the effect of elastic deformations.

The analytical summation for rigid contacts predicts a vanishing friction per particle in the thermodynamic limit ($1/n \to 0$), with a surprisingly low scaling exponent.
This result is general and due to the fact that the pressure vanishes at the edges of the contact.

Because the calculation for the rigid contacts is highly idealised, we also perform molecular-dynamics simulations of a more realistic interface with elasticity and physically implemented pressure distribution.
Our simulations reveal the mechanical boundaries of this regime. 
We have simulated a one-dimensional chain on a surface with a Hertz-like pressure distribution and investigated how the incommensurability, total load, system size, and elasticity affect superlubricity.
The high friction at high load by the transition from incommensurate to commensurate configurations.
At low loads, below the Aubry transition, the friction is smaller and the friction per particle decreases with the system size.
The key is again the vanishing pressure at the edges, which allows the edge atoms to become depinned more easily. In the incommensurate contact with uniform pressure, the friction is dominated by the edge, which under nonuniform pressure becomes more mobile, leading to even lower friction.
Due to long-range relaxtion, the friction moves to a linear scaling as the contact size increases, as is also the case for uniform pressure.
A stiffer slider further enhances this mechanism.

These results demonstrate that nonuniform pressure distributions of realistic contact are not a fundamental barrier to structural superlubricity, as long as the local pressure remains below the Aubry transition.
This raises the hope of new class of structurally superlubric systems that are much more accessible in engineering applications than the more idealised atomically flat contacts often considered in this field.
Sufficiently gently curved interfaces with low bulk elastic constants, covered by flexible layers of 2d materials with relatively high in-plane stiffness.
The low bulk elastic constants are needed to ensure large extended Hertzian contacts, and prevent high local pressures that may harm the 2d material coatings~\cite{Ho2025} while the in-plane stiffness of the 2d material has to be sufficiently high compared to the pressure in the contact to ensure that the system remains safely below the Aubry transition.

\section{Acknowledgments}

This work was funded by the HORIZON-EIC-2021-PATHFINDEROPEN-01 through the project ``SSLiP: Scaling-up Superlubricity into persistence'' (Project No. 101046693).
Views and opinions expressed are however those of the author(s) only and do not necessarily reflect those of the European Union or EIC. Neither the European Union nor the granting authority can be held responsible for them.
The simulations were performed using resources provided Sigma2 - the National Infrastructure for High-Performance Computing and Data Storage in Norway (Project No. NN10020K).

\bibliography{biblio}

\end{document}